# Recoil filters for active correlations method: different scenarios

## Yu.S.Tsyganov

*FLNR, JINR 141980 Dubna, Joliot-Curie str.6, Russian Federation*

tyura@sungns.jinr.ru

### Abstract

*New approach to the life – time estimation for recoil-alpha detected times is presented. Brief description of the **D**ubna **G**as-**F**illed **R**ecoil Separator (**DGFRS**) detection module is presented too. This approach will be used for real-time detecting "active correlations" method with nearest future commissioning of new FLNR (JINR) DC-280 extremely high intense heavy ion cyclotron in the year 2017.*

1. **Introduction**

With commissioning new FLNR (JINR) DC-280 cyclotron in 2017 new requirements are arising in the "active correlation" method [1-4] application to suppress beam associated background signals [5]. One of the is to take into account an influence of background signals to a measured value of live time parameter in a real time mode via detection time-energy-position ER-α correlation chain to stop an actinide target irradiation process for a short time. Some approaches to that problem, to some extent, are reported in [5,6].

In the present paper, some attempt is made to use a more careful method taking into account background signals imitating both alpha and recoil true signals.

2. **Detection module of the DGFRS**

Significant success has recently been achieved in **the** field of SHE synthesis and studies of radioactive properties of superheavy nuclei. With the discovery of the "island of stability" [7] in experiments with $^{48}$Ca projectiles at the DGFRS, one can raise a question about sources and components of such a great event. Intense heavy-ion beams and exotic actinide target materials were certainly strongly required in experiments. However, final products of the DGFRS experiments were rare sequences of decaying nuclei signals. In this connection, **the** role of the DGFRS detection systems was crucial. Specifics of the DGFRS detection system is application of the "active correlations" method [1–4]. Using this technique, it has become possible to provide deep suppression of background products with negligible losses in the value of **the** whole experimental efficiency. Moreover, experiments at the DGFRS, when the above-mentioned method was not applied, yielded ambiguous results [8]. To briefly clarify method application, a process block diagram is shown in Figure 1a-c. A short beam stop was generated by the EVR-α sequence detected in real-time mode.

Note that in most of the DGFRS experiments one of the two first alpha particle signals was used as a trigger signal for a break point in target irradiation.

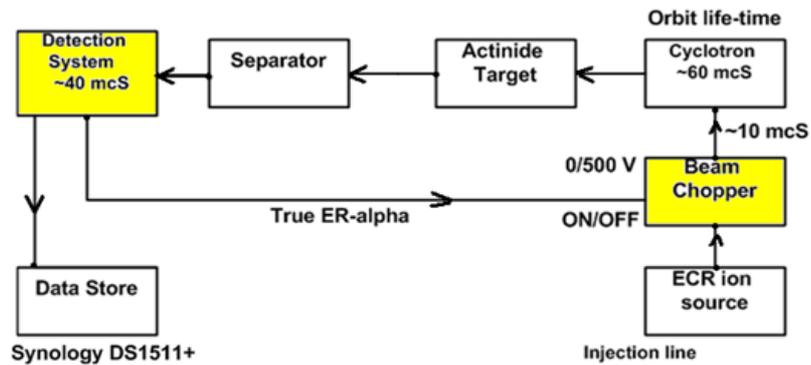

Fig.1a. Schematics of the real-time process to search for ER-α sequence. Parameter 60 µs is corresponded approximately to the orbit life time for $^{48}$Ca ion in main FLNR U-400 cyclotron, 40 µs – dead time of CAMAC electronics.

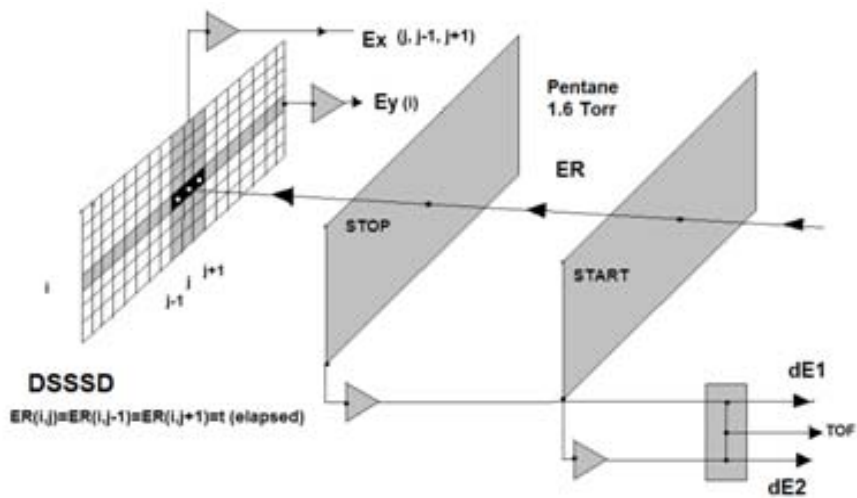

Fig.1b. The DGFRS detection module (schematically). Proportional chambers "start"/"stop", and focal plane *D*ouble *S*ide *S*ilicon *S*trip *D*etector (48x128 strips, Micron Semiconductors, thickness ~300 µm, full depletion) are shown. Low pressure pentane filled ( vapor pressure 1.6 Torr ) gaseous detector creates three signals: TOF (tome-of-flight and two ΔE signals with limited proportionality). An implantation path of ER from a target onto DSSSD detector is shown by an arrow too.

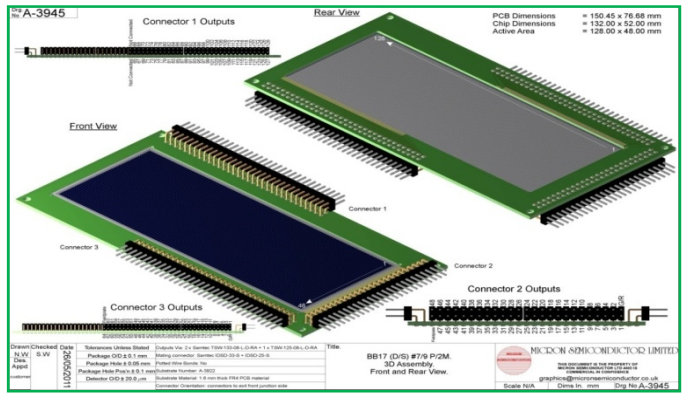

Fig.1c  Micron Semiconductor 48x128 strips detector

1. **An exponential weight function #1**

In this case the a weigh function can be written as: $w_i = e^{-\frac{t_i}{\tau}}$.
As it was shown in [5] the equation for an optimal life-time parameter for beam stop will be:

$$\tau = \frac{w_1 t_1 + w_2 t_2}{w_1 + w_2} = \frac{t_1 e^{-\frac{t_1}{\tau}} + t_2 e^{-\frac{t_2}{\tau}}}{e^{-\frac{t_1}{\tau}} + e^{-\frac{t_2}{\tau}}}.$$

In the Ref.[5] it was shown that few microseconds does it takes for solution with 2-4% precision or even less if one take mean geometry value parameter $\tau_0 = \sqrt[2]{t_1 t_2}$ a first approximation. Note, that both Newton and simple iterations methods gave nearly the same timing results for 3GHz CPU of PC (Fig.2).

Case of "integral filter" $w_i = 1 - e^{-\frac{t_i}{\tau}}$, was considered in [5] too. Of course, an influence of background signals may be taken into account (see below). And it is easy to extend the above approach to the case of n-recoils.

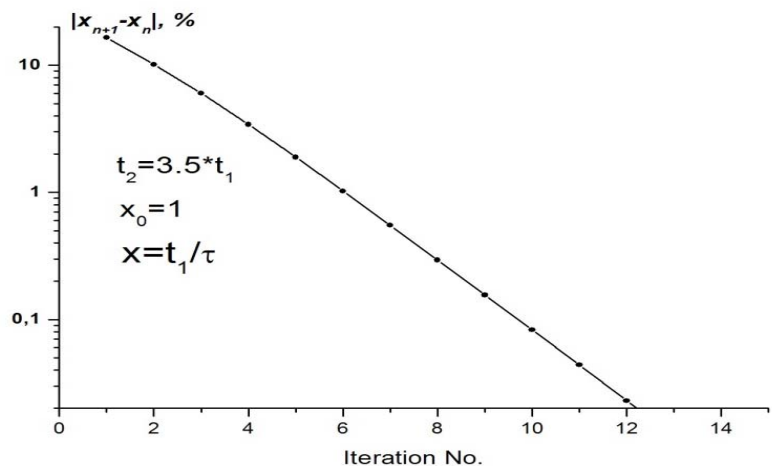

Fig.2 Typical rate of iteration process ( Newton method)

2. **Weight function #2**

Let us consider function like:

$$w(t_i, \tau) = F(bcg) \cdot e^{-\frac{t_i}{\tau}} \cdot (1 - e^{-\frac{t_i}{\tau}}).$$

In that equation [5] it taken into account, in the form of empirical fact, that both factors to register a signal in the o(t$_i$) and in the interval (0,t$_i$) are considered to be the same weight.

Function F(t,τ) may be considered as a factor for the measured ER-α chain to be non-random. For the Poisson like process probabilities for true recoil to correlate with one or more random signals imitating alpha decay and for true α to correlate with one or more recoil imitating signals are

$1 - e^{-\frac{t}{<\tau_\alpha>}}$ and $1 - e^{-\frac{t}{<\tau_{ER}>}}$, respectively.

Therefore,

$$F(bcg) \approx \left(1 - (1 - e^{-\frac{t}{<\tau_\alpha>}})\right) \cdot (1 - (1 - e^{-\frac{t}{<\tau_\alpha>}})).$$

If one introduces a parameter of effective background signal time τ$_{eff}$ as

$$\frac{1}{\tau_{eff}} = \frac{1}{<\tau_\alpha>} + \frac{1}{<\tau_{ER}>}.$$

The resulting equation can be rewritten in the form as:

$$w(t_i, \tau) = e^{-\frac{t_i}{\tau_{eff}}} \cdot e^{-\frac{t_i}{\tau}} \cdot (1 - e^{-\frac{t_i}{\tau}}).$$

Here, t$_i$ – the registered time value (from CAMAC electronics or/and from Windows high precision timer API function) for ER-alpha sequence, <τ$_{α,ER}$> -mean times for alpha and recoil imitator signals, respectively, τ- life time of the nuclide (parameter which have to find).

3. **An alternative (additional) scenario**

An additional way to decrease beam stop number is to use a higher correlation level of ER-α real-time searching procedure. For instance, $ER \cap \alpha_1 \cap \alpha_2$ except for $ER \cap (\alpha_1 \cup \alpha_2)$. As a drawback it will decrease the whole detection efficiency. As concerns to those approaches related with the new FLNR gas-filled separator design, they are outside the scope of this paper. Using a combined mode, when one can in parallel use ER signal to create a shorter beam-off time interval is a reasonable scenario too. The block diagram of the process is shown in the Fig.3.

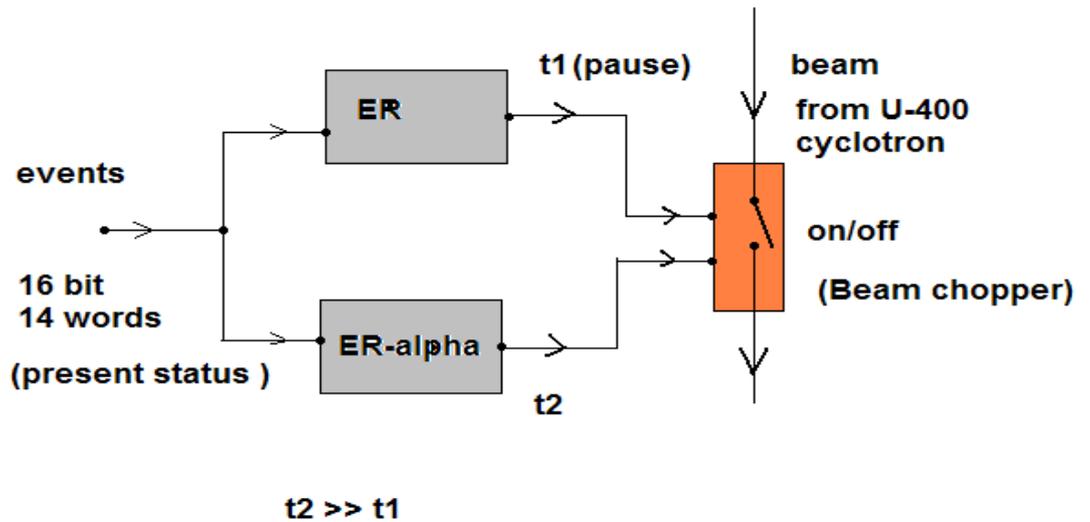

Fig.3 An alternative scenario block diagram; t1 parameter in the figure denotes short interval value after detection of each recoil signal.

## 3. Summary

Different filtering scenarios for non-clear ER-alpha energy-time-position correlated sequences are presented to apply in a nearest future for active correlation technique when DC-280 FLNR, JINR ultra high intense heavy ion beam cyclotron will put into operation.